\documentclass[10pt,journal,compsoc]{IEEEtran}
\usepackage{amsfonts}
\usepackage{amsmath}
\usepackage{amssymb}
\usepackage{graphicx}
\usepackage{epstopdf}
\usepackage{subcaption}
\usepackage{algorithm}
\usepackage{algorithmicx}
\usepackage{algpseudocode}
\floatname{algorithm}{Algorithm}

\ifCLASSOPTIONcompsoc
  \usepackage[nocompress]{cite}
\else
  \usepackage{cite}
\fi
\ifCLASSINFOpdf
\else
\fi
\begin{document}
%
\title{GEMFsim: A Stochastic Simulator for the Generalized Epidemic Modeling Framework}
%
%
%
%

\author{ Faryad Darabi Sahneh, Aram Vajdi$^*$, Heman Shakeri, Futing Fan, Caterina Scoglio
\thanks{This material is based on work supported by the National Science Foundation under Grant No. CIF-1423411.}\thanks{$^\dag$Sahneh, Vajdi, Shakeri, Fan, and Scoglio are with the
Department of Electrical and Computer Engineering at Kansas State University. GEMFsim in MATLAB, R, Python, and C were developed by Darabi Sahneh, Vajdi, Shakeri, and Fan, respectively and supervised by Scoglio.
Emails: \texttt{\{faryad,avajdi,heman,fft,caterina\}@ksu.edu}}}

\IEEEtitleabstractindextext{%
\begin{abstract}
The recently proposed generalized epidemic modeling framework (GEMF) \cite{sahneh2013generalized} lays the groundwork for systematically constructing a broad spectrum of stochastic spreading processes over complex networks. This article builds an algorithm for exact, continuous-time numerical simulation of GEMF-based processes. Moreover the implementation of this algorithm, GEMFsim, is available in popular scientific programming platforms such as MATLAB, R, Python, and C; GEMFsim facilitates simulating stochastic spreading models that fit in GEMF framework. Using these simulations one can examine the accuracy of mean-field-type approximations  that are commonly used for analytical study of spreading processes on complex networks.
\end{abstract}

}

\maketitle

\IEEEdisplaynontitleabstractindextext

%
\IEEEpeerreviewmaketitle

\IEEEraisesectionheading{\section{Introduction}\label{sec:introduction}}

%
%
%
%
\IEEEPARstart{C}{ontagion} phenomena appear in diverse natural and technological contexts, such as infectious disease spreading among humans, computer viruses propagating in computer networks and memes going viral in social networks. In order to understand, predict, and control contagion phenomena, there are models to uncover the underlying mechanisms of spreading processes. Classical models of contagion define some state (or compartment) for the individuals such as \textit{immune}, \textit{susceptible}, \textit{exposed}, \textit{infectious}, \textit{symptomatic}, \textit{recovered}, \textit{dead}, \textit{vaccinated}, and then they define rules for moving from one state to another, assuming the entire population is fully mixed.

During the past two decades, network scientists have demonstrated that interaction among population members can dramatically influence spreading dynamics. Although pioneer works employed random network models, a recent major research direction is to study spreading processes on a generic network with no particular assumption of its structure. In this view, a node represents an individual, links denote interaction among individuals, and a node's current state and the states of its neighboring nodes determine node transitions.

The number of possible spreading models is essentially limitless because the possible node state definitions and rules for node state transitions are not restricted. However, most networked spreading processes share a common fundamental assumption: nodes influence each other through statistically independent pairwise interactions. \textit{Independent} means that the interaction between nodes A and B is statistically independent of the interaction between nodes A and C. \textit{Pairwise} indicates that no higher order interaction is permitted (i.e., joint interaction A--B--C is fully described by A--B, B--C, and A--C interactions)\footnote{Besides the independent pairwise interaction assumption in most epidemic models, other types of interactions exist in the literature. In the contact process \cite{durrett1989contact}, pairwise interactions are exclusive (i.e., a node can only interact with one of its neighbors at a time). In the linear threshold model\cite{acemoglu2011diffusion}, a node interacts with its aggregate neighborhood, and transitions are possible only when the fraction of neighbors in a particular state exceeds a certain threshold value.}.

Based on the independent pairwise interaction characteristic of most spreading models, Sahneh \textit{et al.} \cite{sahneh2013generalized} developed the \textit{generalized epidemic modeling framework}\footnote{In our original article \cite{sahneh2013generalized}, GEMF stands for `\textit{generalized epidemic mean-field}' model. However, because all stochastic descriptions, exact equations, and mean-field equations are detailed in \cite{sahneh2013generalized}, `\textit{generalized epidemic modeling framework}' is a more accurate term, and so it is used in this article.} (GEMF) that facilitates systematic development of a broad spectrum of stochastic spreading processes over complex networks. GEMF is flexible and scalable to incorporate multiple states for the nodes that interact through multiple types of links in a multilayer network structure. In addition to the stochastic description of GEMF-based spreading processes, Sahneh \textit{et al.} have derived the corresponding Kolmogorov equation and the mean-field approximate system of equations. 

In this article, we introduce a simple-to-use tool, GEMFsim, that can numerically simulate any stochastic GEMF-based model. The simulator is based on an adaptation of the Gillespie algorithm \cite{gillespie1976general,gillespie1977exact} to GEMF-based spreading processes in multilayer networks. The Gillespie algorithm can generate a statistically correct trajectory of a continuous-time Markov process. GEMFsim is highly flexible and scalable due to its optimized data structure, and it is capable of simulating spreading processes on networks with millions of nodes. The procedure required for setting up a simulation is simple and systematic: a user inputs the network, transition rules, initial conditions, and stopping criteria. GEMFsim, produced in the Network Science and Engineering (\textsc{NetSE}) research lab at Kansas State University, is available in popular scientific programming platforms such as C, Python, R, and MATLAB; therefore, GEMFsim facilitates simulating stochastic spreading models that fit in GEMF framework. Using these simulations one can examine the accuracy of mean-field-type approximations that are commonly used for analytical study of spreading processes on complex networks.

The rest of the paper is organized as follows. Section \ref{sec:concepts} reviews GEMF, and section \ref{sec:gillespie} proposes a Gillespie-based algorithm for exact simulation of GEMF processes. Section \ref{sec:Applications} concludes the paper by describing several experiments in multiple GEMFsim platforms.

\section{Generalized Epidemic Modeling Framework}\label{sec:concepts}

\subsection{Motivating Example: SIS\ Process on a Graph}\label{sisw}
\begin{figure}[t]
  \centering
    \includegraphics[width=0.3\textwidth]{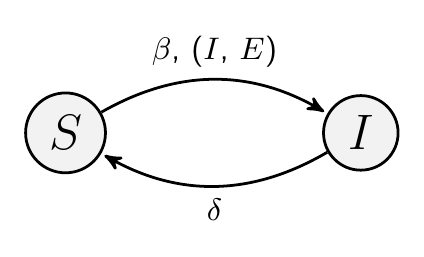}
    \caption{Transition diagram for SIS epidemic model: $\delta$ is the recovery rate of infected nodes, and $\beta$ is the  rate for infecting a susceptible node by an infected neighbor in the network denoted by the edge set $E$.}
\label{sissch}
\end{figure}

Susceptible-infected-susceptible (SIS) is an epidemic model to study infection spreading caused by the interaction of individuals in a network. In the SIS model, individuals are represented by nodes of a graph and possible interactions are the edges of a graph. Node $m$ is a neighbor of node $n$ if it can potentially infect node $n$ directly.
Moreover, the state of node $n$ at time $t$ is denoted by $x_{n}(t)\in\{1,2\}$, where $x_{n}(t)=1$ if the node is susceptible or $x_{n}(t)=2$ if it is infected. In the SIS model, the transition of a susceptible
node to the infected state ($1\rightarrow2$) is result of interaction with an infected neighbor in the network, and the assumption is made that the probability that the node remains susceptible, decays exponentially with the rate $\beta$ as long as the state of the infected neighbor remains unchanged. In addition to the infection process, SIS accounts for the curing process in which an infected node becomes susceptible
($2\rightarrow1$) with a rate $\delta$. The curing process for an infected node is assumed to be independent from the state of its neighbors. Fig. (\ref{sissch}) shows the node-level transitions for the SIS model.\par
According to node-level description of the SIS model, transition of a susceptible node to the infected state depends on states of the neighbors in the network. Hence, mathematical description of the SIS model requires utilization of the network state $X=[x_{1},...,x_{N}]$, which is the joint state of all $N$ nodes in the network. The network state is a continuous-time Markov chain that undergoes transition over a space consisting of $2^{N}$ possible network states. Therefore, the Kolmogorov equation, which governs probability distribution of the Markov process, is a system of $2^{N}$ coupled differential equations which is not computationally tractable for large number of nodes. This necessitates application of approximations \cite{cator2012second,li2012susceptible,van2011n} or simulation in order to study the SIS model. 

\subsection{GEMF Description}
Based on the independent pairwise interaction characteristic of most spreading models, Sahneh \textit{et al.} developed the generalized epidemic modeling framework (GEMF) that facilitates systematic development of a broad spectrum of stochastic spreading processes over complex networks, as comprehensively described in \cite{sahneh2013generalized}.
The SIS model is one of the epidemic models that can be formulated within GEMF. GEMF pertains to an epidemic throughout a network composed of one set of nodes and several layers of contact. We represent the network by  ${\cal G}({\cal V},E_{1},\cdots,E_{L})$, where $L$ is the number of contact layers, ${\cal V}$ is a set of $N$ nodes, and $E_{l}$ is a set of links between the nodes in layer $l$. The incorporation of multilayer topology in GEMF makes it a flexible framework for studying epidemic processes.

Similar to the SIS model, state of node $n$ at time $t$ is a random variable denoted by $x_{n}(t)$. However, each node can assume a node state among $M$ possible states, which are labeled with an integer from $1$ to $M$ (i.e., $x_{n}(t)\in\{1,\cdots,M\}$). 
In GEMF, transitions of $x_{n}$ over the node states are classified in two categories.

\textbf{1. Nodal transitions} of a node are similar to the curing process in the SIS model and they are independent from the states of neighbors in the network. We can generally define nodal transition matrix, $A_{\delta}$, where the element $A_{\delta}(i,j)$
is the transition rate of a node from state $i$ to state $j$. Moreover, 
we can equivalently consider the nodal transition $i\rightarrow j$, as a jump of $x_{n}$ from state $i$ to $j$ with a jump time that is exponentially distributed with the rate $A_{\delta}(i,j)$. In fact, considering $A_{\delta}(i,j)$ as the rate for the jump time gives an insight into our adopted simulation method.

\textbf{2. Edge-based transitions} of a node are analogous to the infecting process in the SIS model. These transitions are caused by interaction with neighbors in the network, and they depend on states of the neighbors. 
 In order to describe edge-based transitions, we can define the transition rate array $A_{\beta}$. Element $A_{\beta}(i,j;l)$ is the rate for transition of a node from state $i$ to $j$ and the transition is result of interaction with a neighbor with state $q(l)$ in layer $l$ . State $q(l)$ is called the influencer state for layer $l$. The influencer state in the SIS model is the infected state that is represented by integer $2$. In GEMF each layer is assumed to have only one influencer state; the layer provides contacts for a node in the influencer state to propagate certain transitions over neighboring nodes. Similar to nodal transitions, we can consider the edge-based transition $i\rightarrow j$ as a jump of $x_{n}$ from state $i$ to $j$ with a jump time that is exponentially distributed with the rate $A_{\beta}(i,j;l)$. However, an edge-based transition of a node is conditioned on the state of the neighboring node; that is the edge-based transition is possible as long as the  neighbor remains in the influencer state. 
\par
Considering the node-level description of transitions in GEMF, the node transition $x_{n}:i\rightarrow j$ may be viable through different possible processes. In general, node $n$ may undergo a transition from state $i$ to $j$ by interacting with neighbors or through a nodal transition. In such a case the processes are assumed to be mutually independent and the transition occurs at a rate that is sum of all rates for the possible processes. That is 
\[%
\begin{array}
[l]{ll}%
\Pr\left(x_{n}\left(t+\Delta t\right)=j\vert\ x_{n}(t)=i\right)=\lambda_{n}\left(i\rightarrow j\right)\Delta t,
\end{array}
\]
where $\lambda_{n}(i\rightarrow j)=r_{1}+\cdots+r_{k}$ and  $r_{1},\cdots, r_{k}$ are rates for the possible processes. 
However, we consider these processes to be competing processes that try to induce the transition $x_{n}:i\rightarrow j$, with jump times distributed as $T_{1}\sim \exp(r_{1}),\cdots ,T_{k}\sim \exp(r_{k})$.
Hence, the actual jump time $T_{x_{n}:i \rightarrow j}$ is the minimum of $\{T_{1},\cdots,T_{k}\}$. Since we assume the competing processes are mutually independent, $T_{x_{n}:i \rightarrow j}$ is distributed exponentially with the rate  $\lambda_{n}(i\rightarrow j)=r_{1}+\cdots+r_{k}$.\footnote{The minimum of  exponentially distributed independent random variables has an exponential distribution with a rate equal to the sum of the individual rates \cite{van2009performance}.} 

Because the edge-based transitions of $x_{n}$ are conditioned on the
state of neighbors in the network, differential equation that governs dynamics of distribution of $x_{n}$ over the node state space $\{1,\cdots,M\}$ is not a closed system. Instead, the joint state of all nodes, defined as $X=[x_{1},...,x_{N}]$, is a continuous-time Markov chain over a space consisting of $M^{N}$ possible network states.  Furthermore, probability distribution of $x_{n}$ over node state space $\{1,\cdots,M\}$ can be obtained as a marginal distribution of X. However, analytical treatment of dynamics of network state, requires solving the Kolmogorov equation for the Markov process, which involves finding transition rates between the network states and solving a system of $M^{N}$ coupled differential equations. The large size of this coupled system, even for a small number of nodes, verifies  
the need for the simulation method presented in this paper.

\section{Event-Based Simulation of GEMF}\label{sec:gillespie}
The network state in GEMF is a Markov process with dynamics that arise from node-level transitions. In this section, we propose an algorithm to sample the Markov process. Furthermore, we discuss GEMFsim, which is a flexible software implementing the proposed algorithm in popular scientific platforms. 
\subsection{Algorithm}

Assuming the joint state of the network at time $t$
is $X(t)=[x_{1},\cdots,x_{N}]$, we can calculate all node-level transition rates $\lambda_{n}(x_{n} \rightarrow j)$ using the nodal transition matrix $A_{\delta}$, edge-based transition array $A_{\beta}$, and the contact network ${\cal G}({\cal V},E_{1},\cdots,E_{L})$, where $\lambda_{n}(x_{n} \rightarrow j)$ is the transition rate of node $n$ from its current state $x_{n}$  to the state $j$.
As described, 
$\lambda_{n}(x_{n} \rightarrow j)$ can be considered to be the rate for the exponential distribution of jump time for the transition, that is  $T_{n}( x_{n} \rightarrow j)\sim \exp(\lambda_{n}(x_{n} \rightarrow j))$. For the network state, occurrence of any node-level jump is a transition of the network state; however, when the first node jumps to another state, transition rates for the other nodes may change.
If we define ${\cal S}$ as the set of all jump times for the node-level transitions,
\[{\cal S}=\{T_{n}( x_{n} \rightarrow j)\vert n\in\{1,\cdots,N\},j\in\{1,\cdots,M\}\},
\] 
then the probability that $T_{n}(x_{n} \rightarrow j)$ would be the minimum of ${\cal S}$ is 
\[%
\Pr\left(T_{n}\left(x_{n} \rightarrow j\right)=\min(S)\right)= \frac{\lambda_{n}(x_{n} \rightarrow j)}{\lambda_{tot}},
\]
where $\lambda_{tot}\triangleq\sum_n\sum_j\lambda_{n}(x_{n} \rightarrow j)$ is the sum of all transition rates corresponding to elements of ${\cal S}$. Using this probability distribution, we can sample one of the node-level transitions that is a transition of the network state. We also must sample the time at which the transition occurs. 
Because elements of ${\cal S}$ have exponential distributions, if  $T=\min({\cal S})$, then $T$ is exponentially distributed with a rate equal to $\lambda_{tot}$. 
\begin{algorithm}
    \caption{\  \   GEMFsim algorithm}
    \begin{algorithmic}[1] 
        \Require $A_{\delta}$, $A_{\beta}$, $W$, $X_{0}$, $q$, \textit{Stop condition}
        \Ensure $event$
       \State $X\leftarrow X_{0}$
       \For{$n=1$ \textbf{to} $N$} 
       \For {$l=1$ \textbf{to} $L$}
       \State $wq(n,l) \leftarrow \sum\limits_{m=1}^{N}W(m,n;l)\delta_{x_{m},q(l)}$
  \EndFor
  \State  $\lambda_{n}\leftarrow \sum\limits_{j=1}^{M}A_{\delta}(x_{n},j)+\sum
  \limits_{l=1}^{L}wq(n,l)A_{\beta}(x_{n},j;l)$   
  \EndFor
  \State $\lambda_{tot}\leftarrow\sum\limits_{n=1}^{N}\lambda_{n}$
  \State $k=0$   
  
     \While{\textit{Stop condition}$=$FALSE}  
     \State  $\alpha\sim \text{Unif}(0,1)$ \Comment{generate $\alpha$ from $\text{Unif}(0,1)$}  
     \State  $\delta t_{k}\leftarrow-\log(\alpha)/\lambda_{tot}$\Comment{time period to the next event}
 \State $P_{1}(n)\leftarrow\lambda_{n}/\lambda_{tot}$
     \State  $n_{k}\sim P_{1}$\Comment{sample $n_{k}$ from probability distribution $P_{1}$}
\State $i_{k}\leftarrow x_{n_{k}}$
 \For{$j=1$ \textbf{to} $M$}
    \State  {\small $\lambda_{n_{k}}(i_{k}\rightarrow j)\leftarrow A_{\delta}(i_{k},j)+\sum
  \limits_{l=1}^{L}wq(n_{k},l)A_{\beta}(i_{k},j;l)$} \EndFor
       \State $P_{2}(j)\leftarrow\lambda_{n_{k}}(i_{k}\rightarrow j)/\lambda_{n_{k}}$
       \State $f_{k}\sim P_{2}$\Comment{sample $f_{k}$ from distribution $P_{2}$}

\State $event(k)\leftarrow(\delta t_{k},n_{k},f_{k},i_{k})$
\State $x_{n_{k}}\leftarrow f_{k}$\Comment{update network state}
\For{$l$ $\mid$ ($q(l)=f_{k}$ \textbf{or} $q(l)=i_{k}$)}\Comment{Update Rates}
\State$\Delta\leftarrow\delta_{q(l),f_{k}}-\delta_{q(l),i_{k}}$
\For{ $n$ $\mid$ $W(n_{k},n;l)\neq0$}
\State{\small $wq(n,l) \leftarrow wq(n,l)+\Delta\times W(n_{k},n;l)$}
\State{\small $\lambda_{n}\leftarrow \lambda_{n}+\Delta\times \sum\limits_{j=1}^{M}W(n_{k},n;l)A_{\beta}(x_{n},j;l)$}
\EndFor
\EndFor
 \State {\small $\lambda_{n_{k}}\leftarrow \sum\limits_{j=1}^{M}A_{\delta}(f_{k},j)+\sum
  \limits_{l=1}^{L}wq(n_{k},l)A_{\beta}(f_{k},j;l)$ }
  \State $\lambda_{tot}\leftarrow\sum\limits_{n=1}^{N}\lambda_{n}$
\State \textit{Update Stop condition}
\State{$k\leftarrow k+1$}
\EndWhile 
        
    \end{algorithmic}\label{alg}
\end{algorithm}

Thus, using the distribution of $T$, we sample a time for the network state transition. The memoryless property of Markov processes allows the entire described procedure to be repeated after the network state is updated. Particularly,
we can directly update the transition rates by the adjustment required due to the change in the state of node $n$ that made the transition, including updating the transition rates of node $n$ and neighbors that can be affected by node $n$. The other rates remain constant.\par
The described simulation method is summarized in Algorithm \ref{alg}, which include the assumption that links in the network can be directed and weighted. If a link is directed from node $m$ to node $n$, node $m$ can induce edge-based transitions on node $n$, but node $n$ cannot induce edge-based transitions on node $m$. Moreover, we can assign a weight to each link  in order to quantify effect of neighbors on edge-based transitions of a node. Rates of edge-based transitions induced by a link are multiplied by weight of the link. In Algorithm \ref{alg}, 
$W(m,n;l)$ is weight of the link directed from node $m$ to node $n$ in layer $l$ of the network, and $W(m,n;l)=0$ indicates no such link. However, implementation of Algorithm \ref{alg} requires that we only store nonzero weights corresponding to links directed from each node. We explain the data structure in more detail in the Appendix.
In Algorithm \ref{alg}, input $q(l)$ is the influencer node state for layer $l$, $A_{\delta}$ and $A_{\beta}$ are nodal transition rates and edge-based transition rates, respectively, and $X_{0}$ is the initial network state. In order to generate a realization of the Markov process over the network state space, we can choose any of the possible node-level transitions according to the probability that the transition occurs, which is $\lambda_{n}(x_{n} \rightarrow j)/\lambda_{tot}$.
Assuming the current network state $X=[x_{1},\cdots,x_{n}]$, node-level transition rates can be calculated as 
\[
\begin{split}
\lambda_{n}(x_{n}\rightarrow j)&=A_{\delta}(x_{n},j)\\
&+\sum_{l=1}^{L}A_{\beta}(x_{n},j;l)\sum_{m=1}^{N}W(m,n;l)\delta_{x_{m},q(l)},
\end{split}
\]
where $\delta_{s,t}$ is Kronecker delta. Moreover, $\lambda_{tot}=\sum_{n=1}^{N}\lambda_{n}$, where $\lambda_{n}=\sum_{j=1}^{M}\lambda_{n}(x_{n}\rightarrow j)$.
In Algorithm \ref{alg} we sample a node-level transition in two steps. First, we 
select node $n$ that will make a transition according to probability
distribution $\Pr(n)=\lambda_{n}/\lambda_{tot}$. After the node
is picked, we select a new node state $j$ according to probability
distribution $\Pr(j\mid n)=\lambda_{n}(x_{n}\rightarrow j)/
\lambda_{n}$. The event-based algorithm explained above is an adaptation of the Gillespie algorithm, which originally was developed for well mixed particles \cite{gillespie1976general,gillespie1977exact}, to GEMF-based processes.

\subsection{GEMFsim Performance}
A broad spectrum of epidemic models can be formulated in the GEMF framework. Hence, GEMFsim is flexible platform capable of simulating various stochastic spreading models.

\subsubsection{Comparison to exact Kolmogorov equations}
\begin{figure}[t]
	\begin{subfigure}{3.4in}
		\includegraphics[width=1\textwidth]{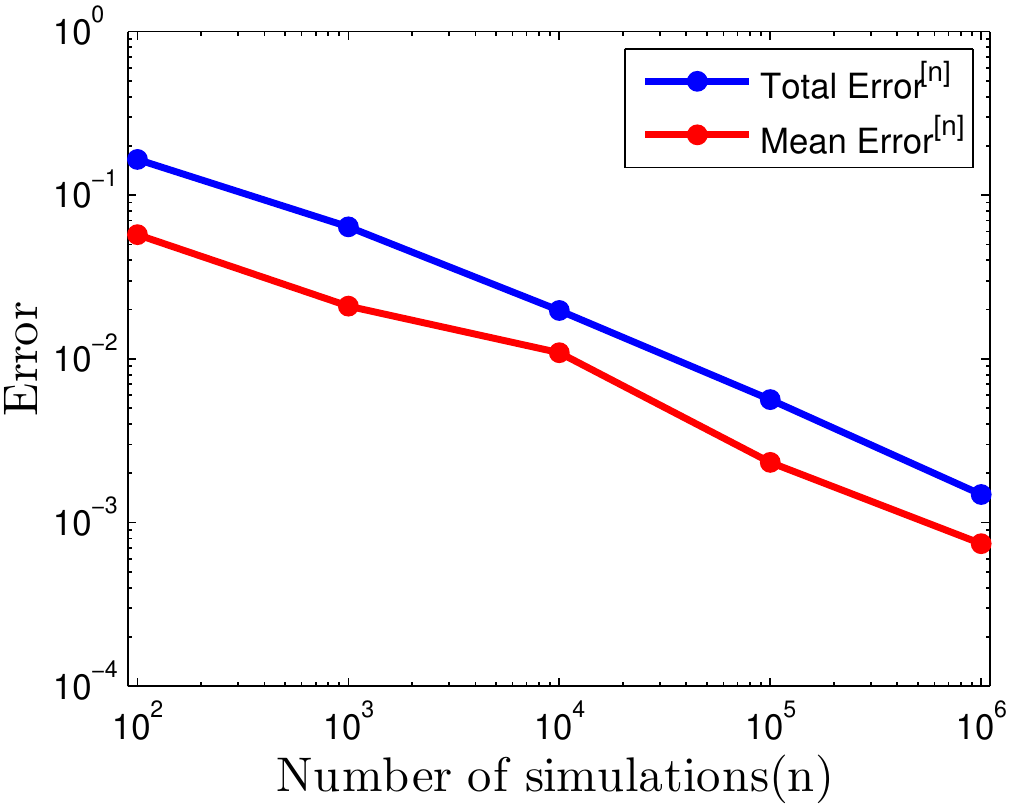}
		\caption{}
		\label{compe}%
	\end{subfigure} 
	\begin{subfigure}{3.4in}
		\includegraphics[width=1\textwidth]{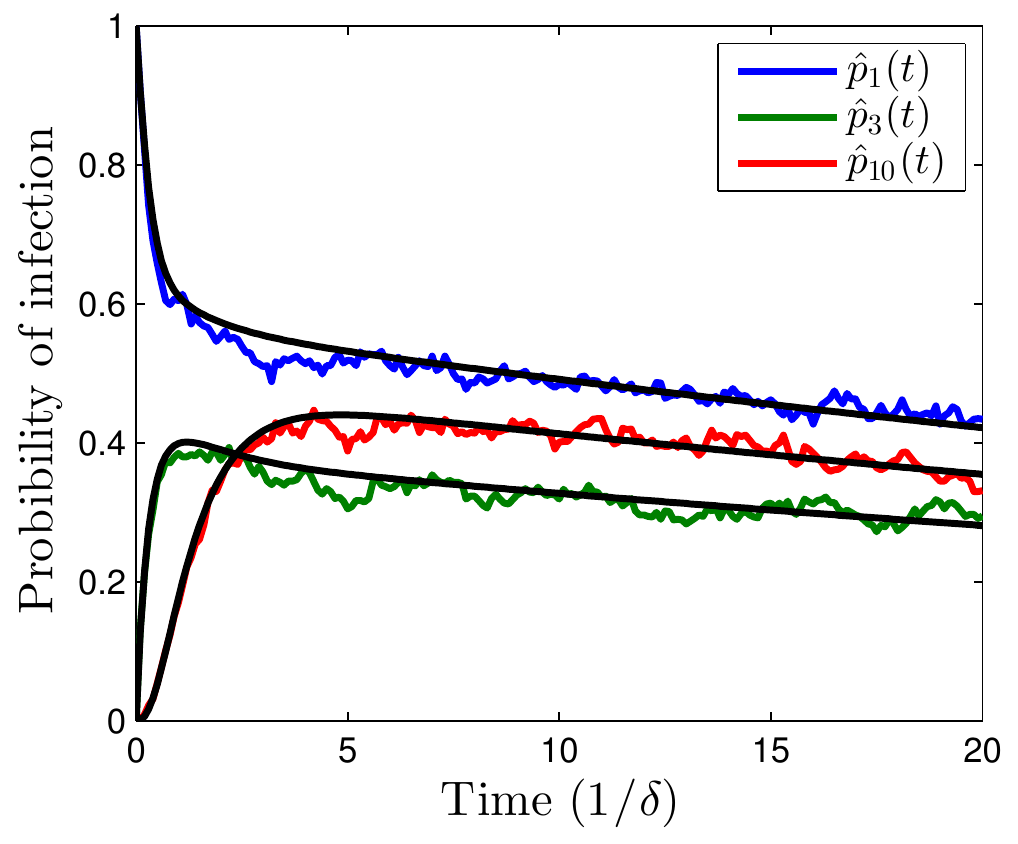}
		\caption{}
		\label{compb}%
	\end{subfigure}
	\caption{The infection probability for each node in a toy network of ten nodes estimated using Monte Carlo simulation in comparison to the exact probability obtained by solving the Kolmogorov equation for the SIS model: (a) total error and mean error defined in Eqs. (1), (2), (b) estimation of infection probability for some nodes obtained by averaging over 1000 simulations. The black (smooth) curves are exact probabilities obtained by solving the Kolmogorov equation.} 
	\label{comp}%
\end{figure}

The event-based algorithm generates exact, statistically correct samples for GEMF processes\cite{gillespie1977exact}. In fact, generated samples for the network state follow a distribution that is a solution of the Kolmogorov equation for the Markov process. In order to experimentally test the distribution of generated samples, we compared results of the Monte Carlo simulation to the exact solution of the Kolmogorov equations for the SIS model. The Kolmogorov equation for the SIS process is a linear system of $2^N$ coupled equations and the size of linear system becomes gigantic, even for moderate 
values of $N$. Therefore, we considered a small network of $N=10$ nodes with SIS parameters of $\delta=1$ and $\beta=2$. Assuming an initial condition in which only one node was infected, we solved the Kolmogorov equation, $\dot{P}(t)=-Q^TP(t)$ \cite[Supplemental Material]{sahneh2013generalized}, where $P$ is a probability distribution over a space consisting of $2^{10}=1,024$ network states and $Q$ is the infinitesimal generator matrix. We then extracted the infection probability of each node, $p^{exact}_i(t)$, as a marginal distribution of $P(t)$. Using Algorithm \ref{alg}, we generated $n$ realizations of the SIS process and obtained an estimation for the infection probability of node $i$, $\hat{p}_i^{[n]}(t)$, as the fraction of realizations when node $i$ was infected at time $t$. Our objective was to observe if the difference between $\hat{p}_i^{[n]}(t)$ and $p^{exact}_i(t)$ decreases as the number of realization $n$ increases. Therefore, we defined two measures of error as
\begin{align}
\textit{Total Error}^{[n]}\triangleq \max_i \max_t |\hat{p}_i^{[n]}(t)-p^{exact}_i(t)|, \\
\textit{Mean Error}^{[n]}\triangleq \max_t \frac{1}{N}\vert\sum_{i=1}^{N}\hat{p}_i^{[n]}(t)-p^{exact}_i(t)\vert.
\end{align}
Fig. (\ref{compe}) shows how the defined measures decreased when the number of  realization $n$ increased.
\subsubsection{Simulation run-time of large networks}
We tested simulation run-time per event for the SIS spreading model on a family of random geometric (RG) networks. In a two-dimensional (2D) model, $N$ nodes are randomly and independently placed in the 2D closed square $[0,\ 1]^2$, and then two nodes are connected to each other via a link if the Euclidean distance between them is less than $r_c$. Algorithm \ref{alg} consisted primarily of two parts, sampling a transition and updating the rates. Our objective was to determine how the run-time for sampling a network state transition changes if the number of nodes increase. Therefor, we generated four RG networks with $N_1=10^3,\ N_2=10^4,\ N_3=10^5,\ N_4=10^6$. In order to maintain the update time in the simulations almost constant, we chose the distance parameters $r_{ck}=\sqrt{N_1/N_k}r_{c1}$ for $k=2,3,4$. Using these values for distance parameters, the average node degrees for all generated networks were the same; this in turn kept the update time constant because, according to Algorithm \ref{alg}, we only update rates for the node that makes the transition and that node's neighbors. For $N_{1}$, we selected $r_{c1}=\sqrt{2\log(N_1)/(\pi N_1})$, in order to guarantee connectivity of generated RG networks \cite{appel2002connectivity}.

\begin{figure}[t]
  \centering
    \includegraphics[width=0.47\textwidth]{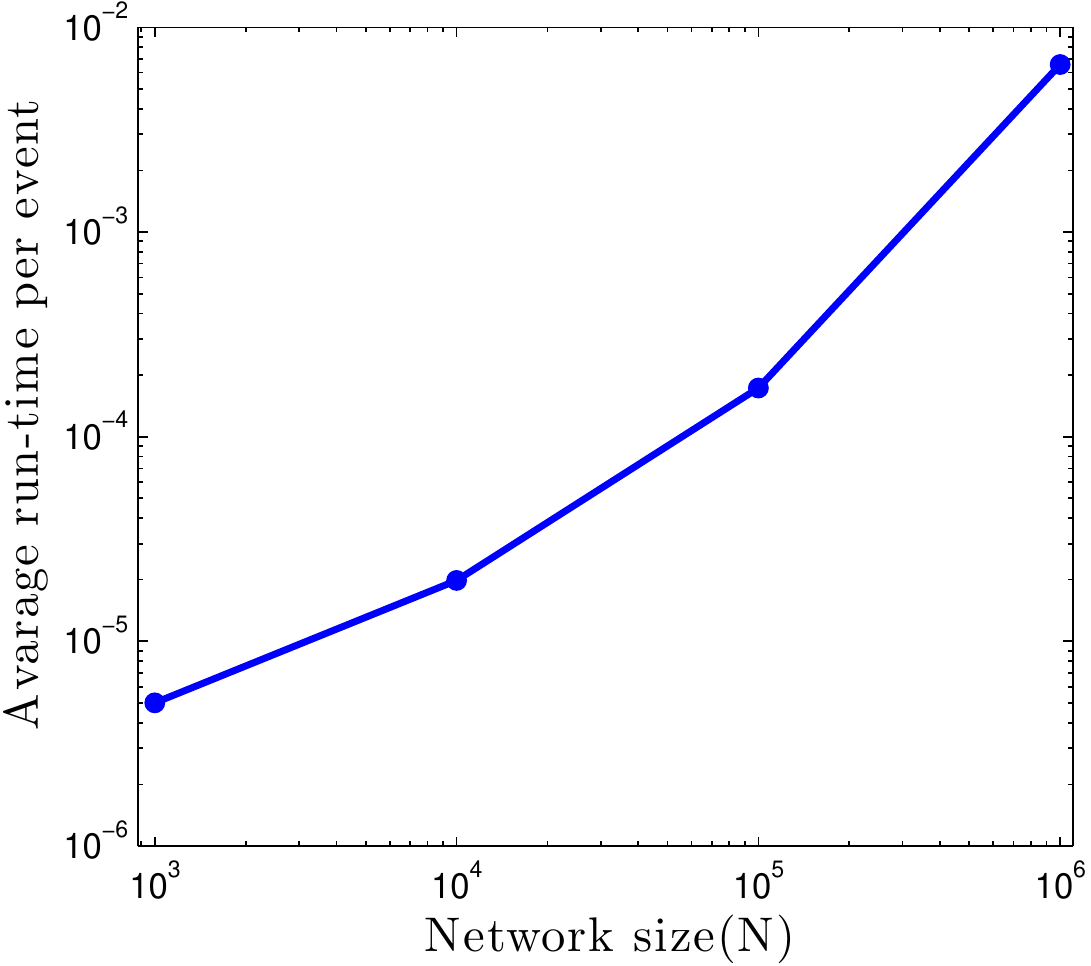}
    \caption{Avarage time to simulate an event in the SIS model, where the contact network is assumed to be a geometric network. The average node degree was kept constant for various network sizes.}
\label{runtime}%
\end{figure}

We simulated the SIS model with parameters  $\delta=1$ and $\beta=3\delta/\rho(G_1)$, where $\rho(G_1)$ is the spectral radius of RG network $G_1$. In Fig. (\ref{runtime}) the average run-time per event is plotted as a function of network size. We performed the simulations via the implementation of GEMFsim in C \cite{GEMFTool} language and it was executed using a machine with 60.0 GB RAM
and two processors of Intel(R) Xeon(R) CPU X5650 @2.67 GHZ .
\section{GEMFsim applications}\label{sec:Applications}
In this section, we show how GEMFsim (implementation of Algorithm \ref{alg}) can be applied to study various compartment models that fit the description of GEMF processes. GEMFsim provides realizations of Markov processes over a space consisting of network states. In theory, GEMFsim can be used to generate enough samples to extract statistics of interest. In fact, any statistics defined in terms of marginal distributions of Markov processes can be estimated using samples generated by the GEMFsim tool. In Section \ref{spatial}
we use this tool to estimate probability distribution of each node in the network over the node states as a function of time. The GEMFsim tool can also be used to estimate the expected population of each node state as described in Section \ref{popu}. These two estimated measures are examples of marginal distributions of Markov processes. However, certain applications of GEMFsim are related to estimating measures that involve joint distribution of individuals. 
Theses are the measures that cannot be approximated using mean-field-type equations. In Section \ref{sirjo}, we use GEMFsim to estimate distributions for extinction time and the fraction of affected individuals in an susceptible-infected-removed (SIR) epidemic. One strength of GEMFsim is its flexibility that enables users to implement complex epidemic simulations. In Section \ref{hemin} we  use GEMFsim to study a complex epidemic scenario that involves competitive epidemic spreading.  
Other applications of GEMFsim are presented in \cite{ebolaNETSE, shakeri2015optimal, shakeri2016generalized}.
\subsection{Simulating the SIS Model}\label{spatial}
\begin{figure}[h!]
	\begin{subfigure}{3.4in}
		\includegraphics[width=1\textwidth]{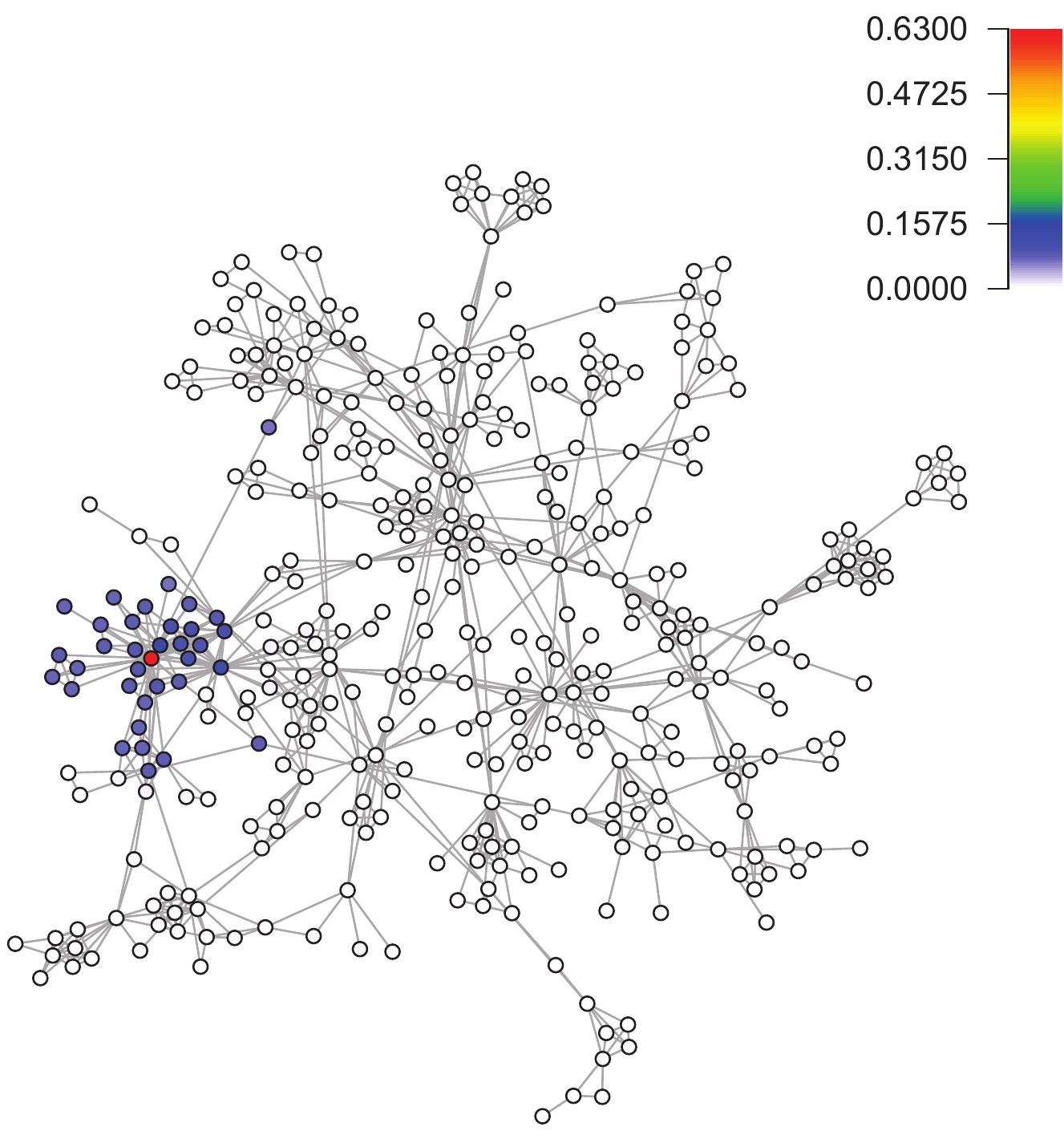}
		\caption{}
		\label{arba}%
	\end{subfigure} 
	\begin{subfigure}{3.4in}
		\includegraphics[width=1\textwidth]{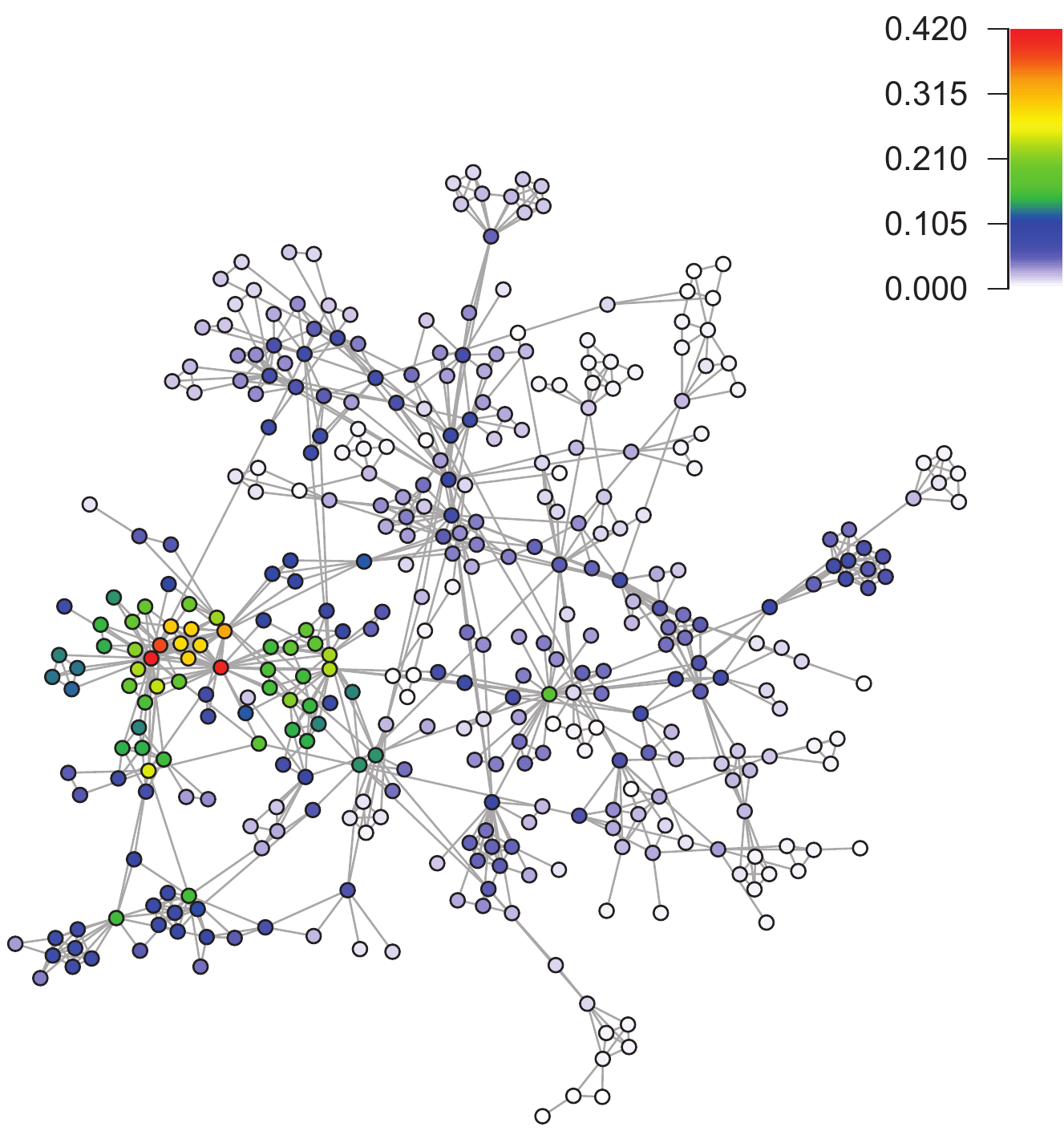}
		\caption{}
		\label{arbb}%
	\end{subfigure}
	\caption{ Result form simulation of SIS spreading over a network. Color of each node represents probability of being infected for the node. (a) probability of being infected at time point $t=0.5$ ($1/\delta$),  (b) probability at time point $t=90$ ($1/\delta$). At $t=0$ only the node with the highest degree was infected. These graphs show evolution of infection in the network }
	\label{arb1}%
\end{figure}

The SIS model, explained in Section \ref{sisw}, is one of the simplest models that can be simulated using GEMFsim. In this model each node is either susceptible (S) or infected (I), as represented by the integers 1 or 2, respectively. If a node is infected, it transmits infection to the susceptible neighbors at a rate $\beta$, and the infected node recovers with the rate $\delta$. We simulated SIS spreading over 
a contact network consisting of one layer of contact ${\cal G}({\cal V},E)$. The network we used was the largest component of the coauthorship network  presented in \cite{newman2006finding}. We assumed that links were undirected and had identical weight. Based on the description of the nodal transition matrix, $A_{\delta}$, and the edge-based transition array, $A_{\beta}$, nonzero elements of them in the SIS model are  $A_{\delta}(2,1)=\delta$ and $A_{\beta}(1,2;1)=\beta$. Moreover, the influencer node state for this model is the infected state (i.e., $q(1)=2$).
Using the implementation of GEMFsim algorithim in R \cite{GEMFTool} we generated 8000 realizations of SIS spreading. We used the results of these simulations to estimate the probability of being infected for each node in the network at various time points. The probability of being infected was estimated as the fraction of SIS realizations in which the node was infected at the given time point. Results for two times are plotted in Fig. (\ref{arb1}). We assumed $\beta=0.23$ and $\delta=1$. The only node that was initially infected in all realizations was the node with the highest degree. 

\subsection{Simulating SIR Model}\label{sirjo}
\begin{figure}[h!]
  \centering
    \includegraphics[width=0.4\textwidth]{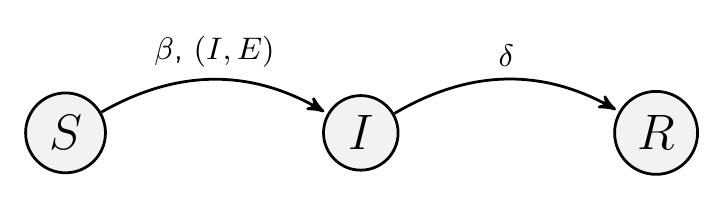}
    \caption{Schematic of node-level transitions in the SIR model}\label{sim:Sir}
\end{figure}
\begin{figure}[h!]
	\begin{subfigure}{3.4in}
		\includegraphics[width=1\textwidth]{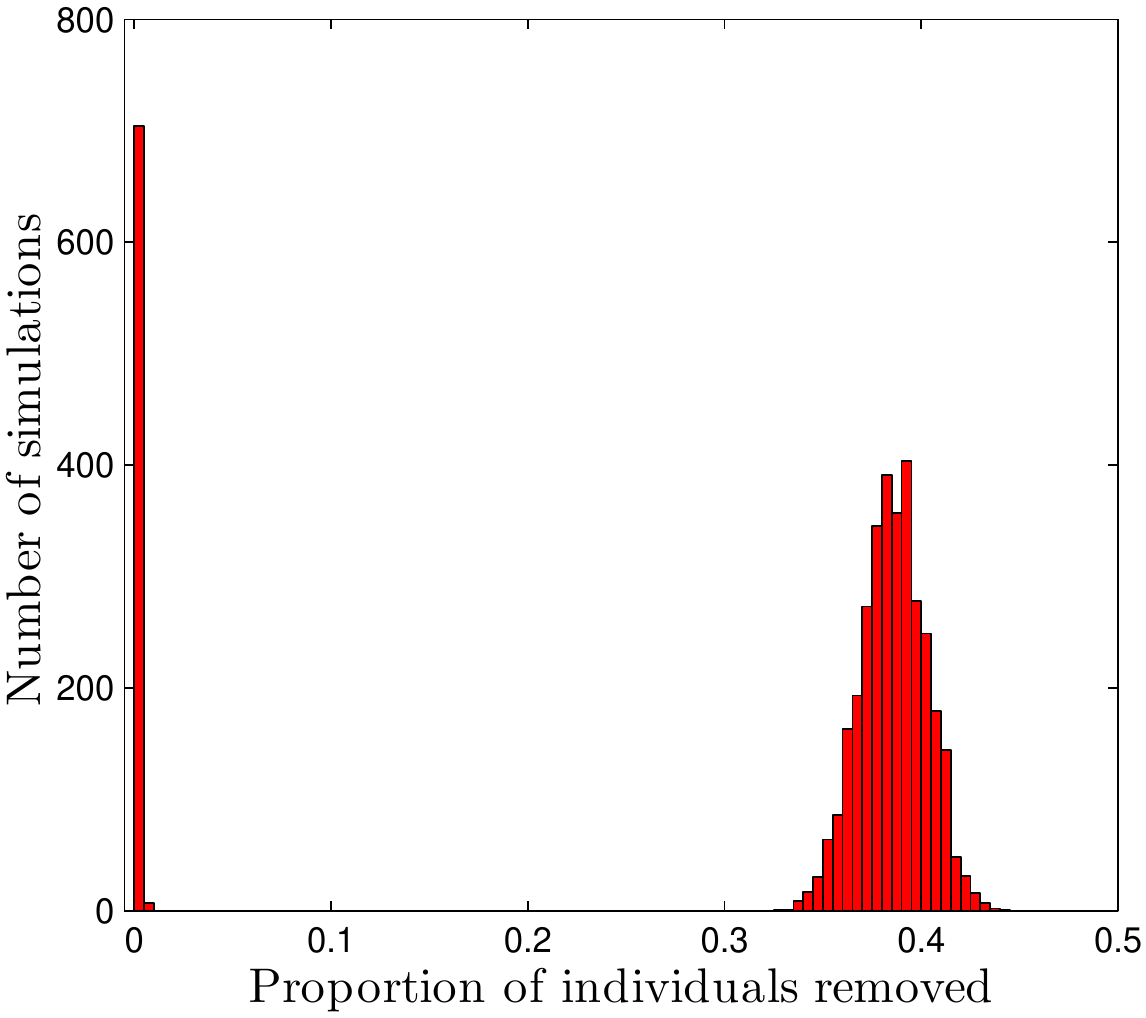}
		\caption{}
		\label{SIRfigR}%
	\end{subfigure} 
	\begin{subfigure}{3.4in}
		\includegraphics[width=1\textwidth]{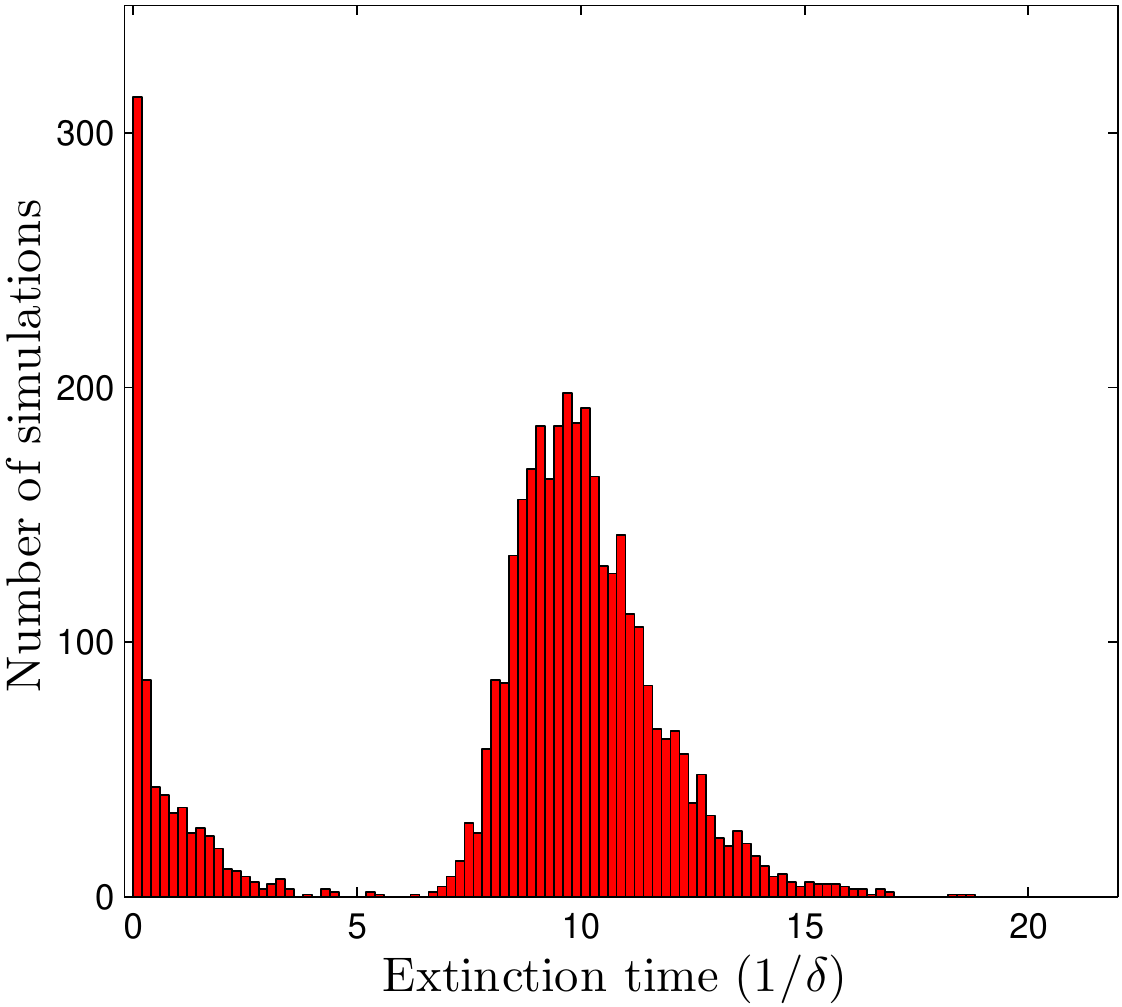}
		\caption{}
		\label{SIRfigT}%
	\end{subfigure}
	\caption{ Results from 4000 realizations of SIR spreading over a network: (a) histogram of the fraction of removed individuals (b) histogram of extinction time defined as the time when the last infected node in the network is removed} 
	\label{SIRfig}%
\end{figure}
In this section we show how GEMFsim can be used to estimate certain statistics which are beyond the scope of mean-field-type approximations. In fact, GEMFsim can be used to generate several realizations of a spreading process and estimate probability distribution for the epidemic measure of interest. We considered an SIR epidemic model in which a susceptible node becomes infected with the rate $\beta $ as a consequence of interacting with an infected neighbor. Moreover, an infected individual transitions to a removed state that may represent the recovered immune state. This transition occurs independently from state of neighbors; in Fig. (\ref{sim:Sir}) the transition rate is shown by $\delta$. In the SIR model, a removed node does not affect its neighbors or undergo any transition and the network eventually reaches an absorbing state in which all individuals are susceptible or removed. Although the time at which the network falls into the absorbing state is not a deterministic variable, the simulation can be used to estimate the probability distribution for the extinction time. The final number of removed individuals is an important measure in epidemiology because it shows the size of outbreak. Similar to extinction time, we can use simulation to estimate probability distribution of  the total number of individuals removed.\par
We Used GEMFsim in MATLAB to generate 4000 realizations of SIR spreading over a directed and weighted Facebook-like social network \cite{opsahl2009clustering,sirnetwork} composed of 1899 nodes and 20296 edges. We assumed initially only the node labeled by integer 1 was infected and the rest of nodes in the network were susceptible. We used transition rates $\beta=0.05$ and $\delta=1$  for the simulation. Node-states in the SIR model are susceptible, infected and removed as labeled by the integers 1, 2, and 3, respectively. The network had one layer of contact with a set of directed and weighted links, and the influencer state was the infected state, was represented by integer 2 (i.e.,  $q(1)=2$). The only nonzero elements of the nodal transition matrix and the edge-based transition array were $A_{\delta}(2,3)=\delta$ and $A_{\beta}(1,2;1)=\beta$. Using simulation we were able to generate histogram of the extinction time and the total fraction of removed individuals in the defined SIR spreading. Fig. (\ref{SIRfig}) shows the total number of affected individuals and extinction time as they follow bimodal distributions.

\subsection{Simulating SAIS Model}\label{popu}

\begin{figure}[t]
  \centering
    \includegraphics[width=0.3\textwidth]{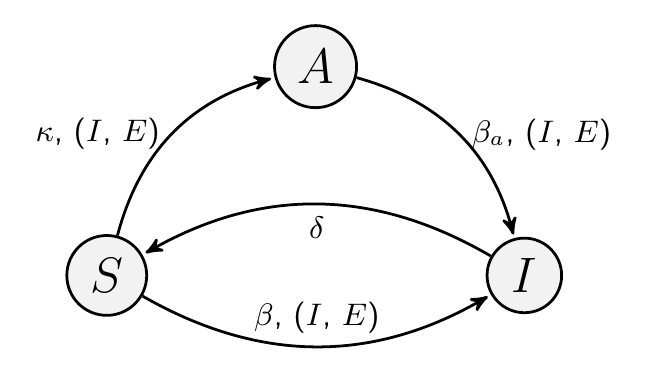}
    \caption{Schematic of node-level transitions in the SAIS model}\label{SAISse}
\end{figure}

A susceptible-alert-infected-susceptible (SAIS) model was developed to incorporate individual reactions to
the spread of a virus \cite{sahneh2011epidemic,sahneh2012existence}. In the SAIS spreading model, each node (individual) is either susceptible (S),
infected (I), or susceptible-alert (A). A susceptible node gets infected with a rate $\beta$ through interaction with an infected node, and an infected node recovers with a rate $\delta$. The SAIS model also accounts for another possibility that a susceptible node can become alert with a rate $\kappa$ if it senses an infected node in its neighborhood. An alert node can also become infected by a process similar to the infection 
process of a susceptible node. However, the infection rate for an alert node, denoted by $\beta_{a}$, is lower due to adoption of preventative behaviors. In order to simulate a realization of the SAIS process, we set up a problem according to the GEMF framework in which three node states (S, A, I) were denoted by integers $1, 2, 3$, respectively.
The network had one layer of contact,  ${\cal G}({\cal V},E)$, where $E$ represents a set of links that could be generally directed and weighted. The influencer state in this model was infected state as represented by integer 3 (i.e.,  $q(1)=3$). The only nonzero element of the nodal transition matrix in the SAIS model is $A_{\delta}(3,1)=\delta$. The nonzero elements of the edge-based transition array are $A_{\beta}(1,3;1)=\beta$, $A_{\beta}(2,3;1)=\beta_{a}$, and $A_{\beta}(1,2;1)=\kappa$. Schematic of node-level transitions in the SAIS model is shown in Fig. (\ref{SAISse})\par
\begin{figure}[t]
  \centering
    \includegraphics[width=0.45\textwidth]{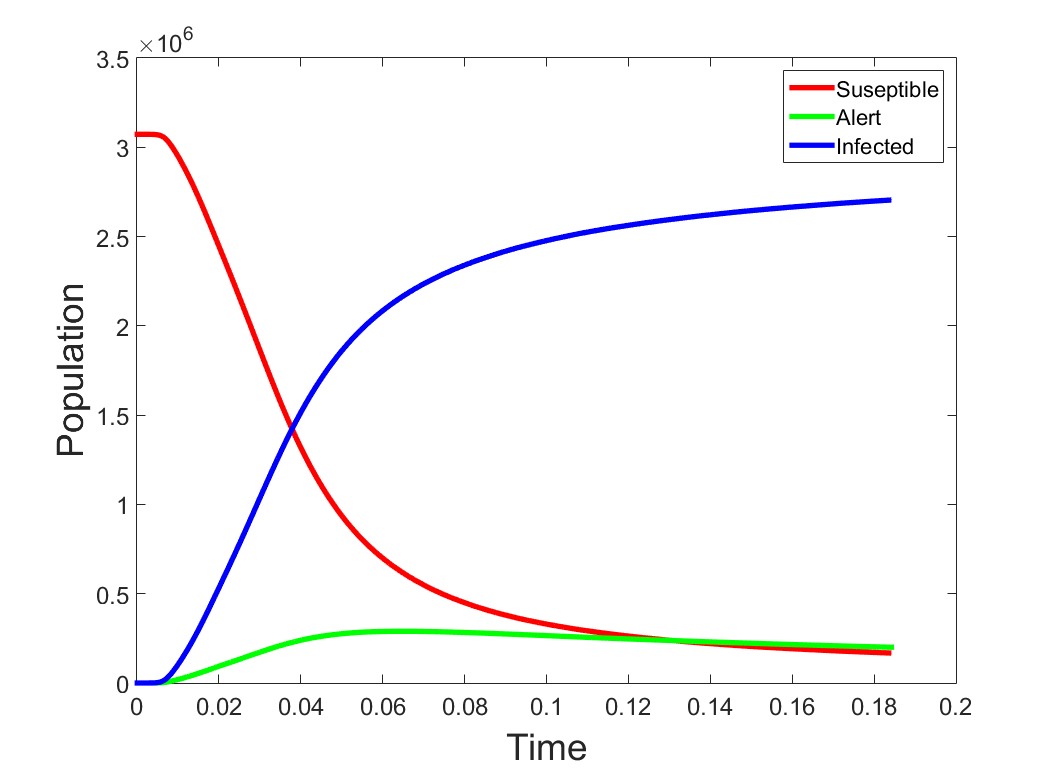}
    \caption{Simulation of SAIS spreading over a large-scale network. Plots represents the population of each node state in the network over time.}\label{SAISsim}
\end{figure}

 
Using implementation of GEMFsim algorithm in C language \cite{GEMFTool} we generated one realization of the SAIS model over a network \cite{DBLP} of 3,072,441 nodes that were connected through 11,7185,083 links. Network links were undirected and had identical weights. Simulation result is shown in Fig. (\ref{SAISsim}). The simulation initially began with 20 infected nodes and 20 nodes in the alert state; the other nodes in the network were initially susceptible. Transition rates for the simulation were $\delta=1,\ \beta=2,\ \beta_{a}=0.4$, and $\kappa=0.2$.

\subsection{Simulating SI$_1$SI$_2$S Model}\label{hemin}
 
\begin{figure}[t]
        \centering
                \includegraphics[width=.5\textwidth]{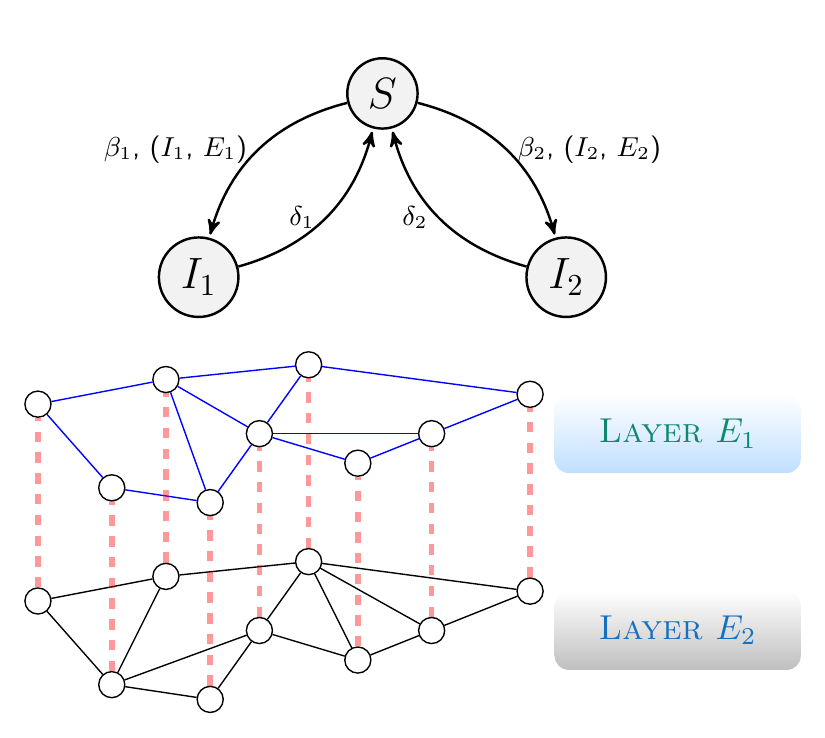}
                        \caption{Node-level transitions in the SI$_1$SI$_2$S spreading model over a two-layer network. Layer $E_{1}$ and $E_{2}$ define two types of contact over the same set of nodes. }\label{fig:SIIS}
\end{figure}

The SI$_1$SI$_2$S model is an extension of the SIS model in which two types of infection can attack a susceptible node \cite{sahneh2014competitive}. However, we assumed a competitive scenario in which the two viruses were exclusive, or a node did not harbor both types of infection simultaneously. Therefore, in this model, each node is either susceptible (S), infected by virus one (I$_1$), or infected by virus two (I$_2$). Similar to the SIS model, infected nodes recover with a rate $\delta_{1}$ or $\delta_{2}$ depending on the infection. In general different infections can be transmitted to a susceptible node through different contacts. In order to account for different means of spreading 
for $I_{1}$ and $I_{2}$, the assumption was made that they spread through different layers of contact such that a susceptible node undergoes a transition to infected state $I_{1}$ ($I_{2}$) with a rate $\beta_{1}$ ($\beta_{2}$) if it is in contact with an $I_{1}$ ($I_{2}$) node through layer $E_{1}$ ($E_{2}$). Fig. (\ref{fig:SIIS}) depicts the SI$_1$SI$_2$S model of spreading. 
The SI$_1$SI$_2$S model can be described in the GEMF framework, by three node states (S, I$_1$, I$_2$) represented by integers $1,2,3$, respectively. The network consists of two layers, ${\cal G}({\cal V},E_{1},E_{2})$, where the first layer spreads $I_{1}$, and the second layer, $I_{2}$. The influencer node state for layer one is $I_{1}$ and the influencer node state for the second layer is $I_{2}$.
The only nonzero elements of nodal transition matrix are $A_{\delta}(2,1)=\delta_{1}$ and $A_{\delta}(3,1)=\delta_{2}$. Nonzero elements of edge-based transition array are $A_{\beta}(1,2;1)=\beta_{1}$ and $A_{\beta}(1,3;2)=\beta_{2}$. In general, $E_{1},E_{2}$ could be two different sets of links between nodes. However, if both types of infection use the same kind of contacts to spread, $E_{1}$ and $E_{2}$ are similar.
\begin{figure}[top]
  \centering
    \includegraphics[width=0.5\textwidth]{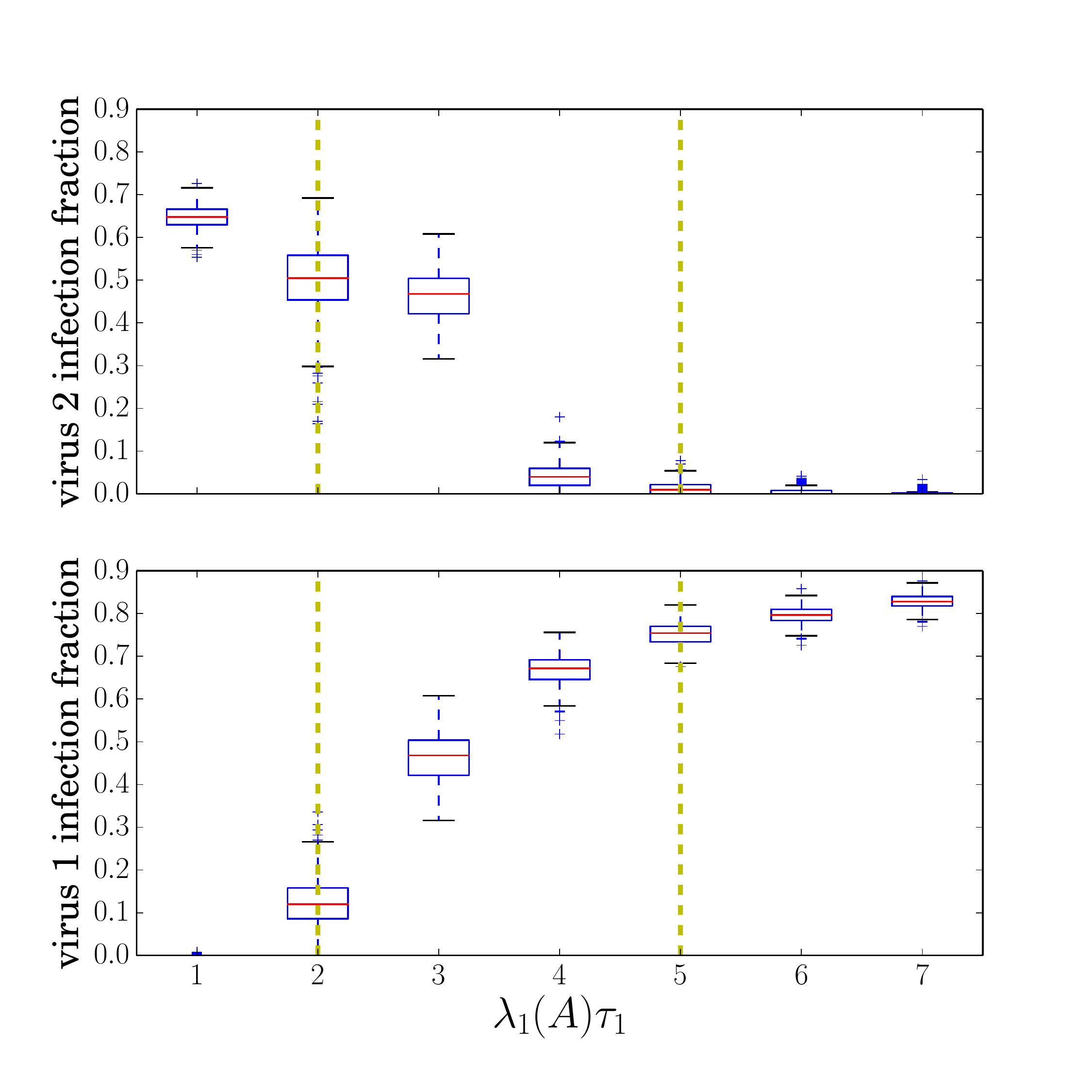}
    \caption{Fraction of nodes infected by virus type $2$ (above) and virus type $1$ (below) in the SI$_1$SI$_2$S competitive spreading model. Infection strength of $I_{2}$, $\tau_{2}$, was $5/\lambda_1(B)$, while the infection strength of $I_{1}$, $\tau_{1}$, varied. If\ \  $2/\lambda_1(A)\leq \tau_{1} \leq5/\lambda_1(A)$,  viruses coexist; only one virus survives outside this region.} \label{sim:SI1I2S}
\end{figure}
The SI$_1$SI$_2$S model described above, exemplifies a competitive spreading scenario in which two types of infection try to invade a network.
However,  mean-field-type approximation showed, in a network with two different layers of contact, $I_{1}$ and $I_{2}$ can coexist depending on their infection rates \cite{sahneh2014competitive}. We used implementation of GEMFsim in Python \cite{GEMFTool} to show this coexistence via simulation.
We adopted a network of 500 nodes with two different contact layers, $E_{1}$ and $E_{2}$. We assumed $I_{1}$ spreads through contact layer $E_{1}$, which is a scale-free network \cite{barabasi1999emergence} of 2,475 edges and that $I_{2}$ uses a geometric network, $E_{2}$, of 3,560 edges to invade the nodes. Assuming $B$ 
($A$) is the adjacency matrix for contact layer $E_{2}$ ($E_{1}$), we used value of  $5/\lambda_1(B)$ as the infection strength $\tau_2 = \beta_{2}/\delta_{2}$ in which $\lambda_1(B)$ is the largest eigenvalue of adjacency matrix $B$. However, for infection strength $\tau_1 = \beta_{1}/\delta_{1}$
we used seven values from $\tau_1 =1/\lambda_1(A)$ to $\tau_1 =7/\lambda_1(A)$; for each value of $\tau_{1}$ we generated $500$ realizations of SI$_1$SI$_2$S processes. For all simulations we assumed that each virus had initially infected $2\%$ of the nodes. Fig. (\ref{sim:SI1I2S}) shows metastable state population sizes extracted from simulations for values of $\tau_{1}$. As shown in the figure, either one of the viruses prevails or the viruses coexist depending on the value of $\tau_{1}$.

\section{Conclusion}
Networked spreading processes have attracted substantial interest in the dynamical systems and controls community; possibly mainly because the process essentially belongs to the notion of networked Markov processes, or more generally, networked dynamical systems, and it has functional application in sociotechnological systems. However, exact analytical study of networked spreading processes is extremely difficult, if not impossible, due to the gigantic state-space size of possible network states. For example, Kolmogorov equations for stochastic GEMF-based processes have a state-space size of $M^N$, which exponentially grows by the network size. However, utilization of moment-closure techniques in which higher order moments are approximated by lower order moments have facilitated analytical study of networked spreading processes. For example, a first order moment-closure technique, also referred to as mean-field-type approximation, leads\footnote{For a rigorous step-by-step development of mean-field equations for GEMF-based processes, please refer to \cite{sahneh2013generalized}.} to a nonlinear system of differential equations with $MN$ states, which linearly grows by $N$. Mean-field approximate equations have two major shortcomings that limit the equations' applicability. First, mean-field equations can be significantly inaccurate for certain networks and certain parameter spaces; hence questioning results obtained solely based on them. Second, to our knowledge, no rigorous result has yet quantified the extent of inaccuracies. Despite these concerns, there are not much alternatives to mean-field equations for analytical studies of this highly complex problem. In order to have reliable results, mean-field results should always be tested against actual exact numerical solutions. In this way, mean-field equations merely serve as a guide to understanding and controlling spreading processes. We believe that GEMFsim, which is implemented in popular scientific programming languages MATLAB, R, Python, and C, provide opportunities for growing research on spreading processes.

Future research directions to improve GEMFsim can include implementing GEMFsim for parallel processing, especially with graphic processing units (GPU), in order to enable rapid simulation of even larger networks. Another important
direction is developing a $\tau$--leap method \cite{gillespie2001approximate}, which can dramatically accelerate simulation runtime while generating acceptable inaccuracy for large network sizes.
\appendix
\subsection*{Data Structure}
We implemented the GEMFsim algorithm and is available in popular scientific programming platforms such as C, Python, R, and Matlab. Prior to using any of these implementations, however, a user must provide a set of the following input parameters to describe the epidemic model:
\begin{itemize}
\item \texttt{N} is the number of nodes in the network.
\item \texttt{M} is the number of compartments (node state) that a node can assume. These compartments are labeled with integers from \texttt{1} to \texttt{M}.
\item $\texttt{A}_{d}$ is the nodal transition matrix with \texttt{M}$\times$\texttt{M} dimensions. The element $\texttt{A}_d\texttt{(i,j)}$ specifies the nodal transition rate from compartment \texttt{i} to \texttt{j}.
\item \texttt{L} is the number of layers that comprise the interaction network. Each layer is labeled with an integer between  \texttt{1} to \texttt{L}.
\item \texttt{q} is a vector of dimension \texttt{L} that stores the influencer compartments. Element \texttt{q(l)} is the influencer compartment corresponding to layer \texttt{l}.
\item $\texttt{A}_b$ is the edge-based transition array with \texttt{M}$\times$\texttt{M}$\times$\texttt{L} dimensions. Element $\texttt{A}_b\texttt{(i,j,l)}$ is the transition rate of a node from compartment $\texttt{i}$ to $\texttt{j}$ if it is  connected to a neighbor in layer \texttt{l} through a link of weight equal to 1 while the state of neighbor is $\texttt{q(l)}$.
\item The three input parameters \texttt{Neighbors}, $\texttt{I}_1$, $\texttt{I}_2$ specify the interaction network.  $\texttt{I}_1$ and $\texttt{I}_2$ are $L\times N$ matrices and   \texttt{Neighbors} is a list that contains $L$ matrices where each matrix has two rows.  The neighbors of node \texttt{n} in layer \texttt{l} are elements of  vector 
$\texttt{v}=\texttt{Neighbors}\{\texttt{l}\}(1,\texttt{I}_1(\texttt{l},\texttt{n}):\texttt{I}_2(\texttt{l},\texttt{n}))$ where $\texttt{I}_1(\texttt{l},\texttt{n}):\texttt{I}_2(\texttt{l},\texttt{n})$ is a sequence of numbers from  $\texttt{I}_1(\texttt{l},\texttt{n})$ to $\texttt{I}_2(\texttt{l},\texttt{n})$,  increasing by an increment of 1. These are the nodes that  can be potentially affected by the node $\texttt{n}$. Moreover, the weight of the link between node $\texttt{n}$ and its neighbors, obtained from vector $\texttt{v}$, are 
$\texttt{Neighbors}\{\texttt{l}\}(2,\texttt{I}_1(\texttt{l},\texttt{n}):\texttt{I}_2(\texttt{l},\texttt{n}))$ respectively.
\item $\texttt{x}_0$ is a vector with length $\texttt{N}$ that stores the initial state of each node. For example, if node $\texttt{n}$  initially is in compartment $\texttt{m}$,  then $\texttt{x}_0(\texttt{n})=\texttt{m}$. \textit{GEMF\_SIM} uses $\texttt{x}_0$ as an initial condition for spreading  simulation.
\item \texttt{stop condition} determines when the simulation should stop. For example, the simulation stops if the number of events reaches a specified number or if the total time for the evolution passes a specified number.
\end{itemize}
After generating one realization of the the Markov process, GEMFsim outputs the summary of simulation in four vectors $\texttt{t}_e$, $\texttt{n}_e$, $\texttt{i}_e$, $\texttt{j}_e$, where $\texttt{t}_e(\texttt{k})$ is the time interval between the events $\texttt{k-1}$  and  $\texttt{k}$;and event $\texttt{k}$ is a transition that node $\texttt{n}_e(\texttt{k})$  changes its state from compartment $\texttt{i}_e(\texttt{k})$  to compartment $\texttt{f}_e(\texttt{k})$ 
\bibliographystyle{IEEEtran}
\bibliography{GEMFsim}

%
%
%
%
%
%
%
%
%
%

%




\end{document}